\documentclass[preprint,showpacs,aps,pra]{revtex4}
\usepackage{epsf}
\usepackage{dcolumn}
\usepackage{bm}
\everymath{\displaystyle}
\begin{document}

\title{Energy expectation values of a particle in nonstationary fields}

\author{Alexander J. Silenko}
\email{alsilenko@mail.ru} \affiliation{Research Institute for
Nuclear Problems, Belarusian State University, Minsk 220030,
Belarus\\
Bogoliubov Laboratory of Theoretical Physics, Joint Institute for
Nuclear Research, Dubna 141980, Russia}

\date{\today}

\begin{abstract}
We show that the origin of the nonequivalence of Hamiltonians in different representations is a change of the form of the time-derivative operator at a time-dependent unitary transformation. This nonequivalence does not lead to an ambiguity of the energy expectation values of a particle in nonstationary fields but assigns the basic representation. It has been explicitly or implicitly supposed in previous investigations that this representation is the Dirac one. We prove the alternative assertion about the basic role of the Foldy-Wouthuysen representation. We also derive the general equation for the energy expectation values in the Dirac representation. As an example, we consider a spin-1/2 particle with anomalous magnetic and electric dipole moments in strong time-dependent electromagnetic fields. We apply the obtained results to a spin-1/2 particle in a plane monochromatic electromagnetic wave and give an example of the exact Foldy-Wouthuysen transformation in the nonstationary case.
\end{abstract}
 \maketitle

\section{Introduction}

The important problem of energy expectation values (EEVs) of a
particle in nonstationary external fields has a long history. The
basic equation describing a unitary transformation of a
time-dependent Hamiltonian operator is well known
\cite{FW,Messiah}. The problem of the EEVs has been considered in
detail in Refs. \cite{Nietoold,Gol,Nieto,Fearing}. In these works, the dependence
of the EEVs on the representation used has been clearly
demonstrated. It has been claimed in Refs.
\cite{Gol,Nieto,Fearing} that this fact definitely results in a
physical nonequivalence of the initial and transformed Hamiltonians in
the time-dependent case. The problem of physical equivalence of these Hamiltonians has been recently
reexamined in Refs. \cite{Arminjon,GorNeznPRD,GorNeznJdP}. This problem is very important in
relation to the Foldy-Wouthuysen (FW) transformation
\cite{FW}.

Gorbatenko and Neznamov \cite{GorNeznPRD,GorNeznJdP} have demonstrated the
possibility of connecting Hamiltonians in different representations and have also considered the problem of their physical equivalence.

Goldman \cite{Gol} and Nieto \cite{Nieto} have
shown that derivation of the EEVs from the time-dependent 
Hamiltonians may lead to controversial and even incorrect results. They proceeded from the nonequivalence of different representations
in the time-dependent case and explicitly or implicitly supposed that the basic representation is the Dirac one. The same supposition was used in Refs. \cite{Fearing,Kupersztych}.

We will show that further developments of the theory of the FW
transformation fulfilled after the publication of Refs.
\cite{Nietoold,Gol,Nieto,Fearing,Kupersztych} lead to a different conclusion about the basic representation. We will also give a first example of the exact FW transformation in the nonstationary case.

We use the system of units with $c=1$ while $\hbar$ is included in quantum-mechanical equations.

\section{Unitary transformations of a time-dependent Hamiltonian operator}\label{unitary}

Operators used in quantum mechanics are self-adjoint. Many authors claim that such operators should be Hermitian. However, this assertion is inexact. When any operator is Hermitian, it does not necessarily mean that this operator is self-adjoint. A densely defined operator $T$ on the Hilbert space $\bm{\mathcal{H}}$ is called \emph{symmetric} (or \emph{Hermitian}) if $T\subset T^\ast$, that is, if $D(T)\subset D(T^\ast)$ and $T\varphi=T^\ast\varphi$ for all $\varphi\in D(T)$. Here $T^\ast$ is the adjoint operator and $D(T^\ast)$ is the domain of its definition. Equivalently, $T$ is symmetric if and only if $(T\varphi,\chi)=(\varphi,T\chi)$ for all $\varphi,\chi\in D(T)$ \cite{footnote}. $T$ is called \emph{self-adjoint} if $T=T^\ast$, that is, if and only if $T$ is symmetric and $D(T)=D(T^\ast)$.

Thus, every self-adjoint operator is symmetric. However, the converse may be unsatisfied. Let the operator $T=i(d/dx)$ be defined on the interval $[0,1]$ as follows:
$$D(T)=\{\varphi|\varphi\in AC[0,1], ~~~ \varphi(0)=\varphi(1)=0\}.$$
It can be proven (see Refs. \cite{Reed-Simon}) that the operator $T$ is closed and symmetric (Hermitian) but it is not self-adjoint.


If $T$ is continuous and is defined on the whole Hilbert space, $D(T)=\bm{\mathcal{H}}$, then the symmetric operator $T$ is also self-adjoint.

A unitary transformation of any operator except for the Hamiltonian one
is given by \begin{equation} A'=UAU^{-1}, \label{eqA}
\end{equation} where $U$ is a unitary operator
transforming the wave function $(\psi'=U\psi)$ from the unprimed
representation to the primed one. The transformation of the
Hamiltonian operator is different because this operator is defined
by
\begin{equation} i\hbar\frac{\partial\psi}{\partial t}={\cal
H}\psi.
\label{eqCorSc}
\end{equation}
As a result, the transformation also involves the operator
$i\hbar\,(\partial/\partial t)$. The transformed Hamiltonian is
given by \cite{FW,Messiah}
\begin{equation} {\cal
H}'=U\left({\cal H}-i\hbar\frac{\partial}{\partial
t}\right)U^{-1}+i\hbar\frac{\partial}{\partial t}=U{\cal
H}U^{-1}-i\hbar U\frac{\partial U^{-1}}{\partial t}. \label{TrHam}
\end{equation}
Since $\partial(UU^{-1})/(\partial t)=0$, the result of the
transformation can also be presented as follows \cite{GorNeznPRD}:
\begin{equation} {\cal
H}'=U{\cal H}U^{-1}+i\hbar \frac{\partial U}{\partial
t}U^{-1}.\label{TrHamtw}
\end{equation} Evidently, the connection between the initial and transformed
Hamiltonians substantially differs from Eq. (\ref{eqA}).

The EEV of the particle is defined by
\begin{equation} E(t)=\int{\psi^\dag(\bm r,t){\cal H}(t)\psi(\bm r,t)dV}. \label{EEV}
\end{equation}

For a particle in nonstationary fields, the operators ${\cal H}$
and $U$ explicitly depend on time. In this case, the EEVs in the
unprimed and primed representations are not equal to each other
\cite{Nietoold,Gol,Nieto,Fearing,Arminjon,Kupersztych}. The use of Eqs. (\ref{TrHam}) and
(\ref{TrHamtw}) results in
\begin{equation} \begin{array}{c} \int{{\psi'}^\dag(\bm r,t){\cal H}'(t)\psi'(\bm r,t)dV}=
\int{{\psi}^\dag(\bm r,t){\cal H}(t)\psi(\bm r,t)dV}\\
-i\hbar\int{{\psi}^\dag(\bm r,t)U\frac{\partial
U^{-1}}{\partial t}\psi(\bm r,t)dV}=\int{{\psi}^\dag(\bm r,t){\cal H}(t)\psi(\bm r,t)dV}\\
+i\hbar\int{{\psi}^\dag(\bm r,t)\frac{\partial U}{\partial t}U^{-1}\psi(\bm r,t)dV}.
\end{array} \label{Hconn}
\end{equation}

A comparison of Eqs. (\ref{EEV}) and (\ref{Hconn}) demonstrates the 
nonequivalence of the initial and
transformed Hamiltonians in the time-dependent case
\cite{Gol,Nieto,Fearing,Arminjon,Kupersztych}.
Equation (\ref{Hconn}) shows that Eq. (\ref{EEV}) for the particle EEV can
be satisfied in one and only one representation. This representation
is basic and it cannot be physically equivalent to others.

It has been claimed by Gorbatenko and Neznamov \cite{GorNeznPRD,GorNeznJdP} that Hamiltonians related to
each other by unitary transformations are physically equivalent. However, the problem of the EEVs was not considered in Refs. \cite{GorNeznPRD,GorNeznJdP}.

Nieto \cite{Nieto} has stated that the operator $U{\cal H}U^{-1}$ has the same expectation values as ${\cal H}$:
$$\int{{\psi'}^\dag(\bm r,t)U{\cal H}(t)U^{-1}\psi'(\bm r,t)dV}=
\int{{\psi}^\dag(\bm r,t){\cal H}(t)\psi(\bm r,t)dV}.$$ Let the unprimed representation be basic
and $U$ is the unitary transformation operator from the unprimed
representation to the primed one.
Therefore, the energy operator in the primed representation is $\widetilde{{\cal H}'}=U{\cal H}U^{-1}$ but not ${\cal H}'$ (see Ref. \cite{Lusanna}).
This property allows us to obtain correct EEVs in any representation. If the Hamiltonian in a nonbasic (primed) representation is known, the EEV is given by
\begin{equation} \begin{array}{c} E(t)=\int{{\psi'}^\dag(\bm r,t)\widetilde{{\cal H}'}(t)\psi'(\bm r,t)dV}=\int{{\psi'}^\dag(\bm r,t){\cal H}'(t)\psi'(\bm r,t)dV}\\
-i\hbar\int{{\psi'}^\dag(\bm r,t)\frac{\partial
U}{\partial t}U^{-1}\psi'(\bm r,t)dV}=\int{{\psi'}^\dag(\bm r,t){\cal H'}(t)\psi'(\bm r,t)dV}\\
+i\hbar\int{{\psi'}^\dag(\bm r,t)U\frac{\partial U^{-1}}{\partial t}\psi'(\bm r,t)dV} \end{array} \label{EEVprim}
\end{equation} or \begin{equation} \begin{array}{c} E(t)=\left\langle\widetilde{{\cal H}'}\right\rangle=\left\langle{\cal H}'\right\rangle-i\hbar\left\langle\frac{\partial
U}{\partial t}U^{-1}\right\rangle=\left\langle{\cal H}'\right\rangle+i\hbar\left\langle U\frac{\partial U^{-1}}{\partial t}\right\rangle. \end{array} \label{EEVprimEq}
\end{equation} The possibility to use any representation for a correct description of a quantum system corresponds to fundamental principles of quantum mechanics (QM).

It is easy to explain the origin of the nonequivalence. Equation (\ref{eqCorSc}) can be transformed to the form
\begin{equation} i\hbar U\frac{\partial}{\partial t}U^{-1}\psi'=i\hbar\left(\frac{\partial}{\partial t}\right)'\psi'=\widetilde{{\cal H}'}\psi'.
\label{eqCorNw}
\end{equation} Thus, time-dependent unitary transformations change the form of the operator $i\hbar(\partial/\partial t)$ (as well as that of the time operator, $t$). The spatial components of the four-momentum operators $p_\mu=i\hbar(\partial/\partial x^\mu)$ and $x^\mu$ possess similar properties. Therefore, the operator $i\hbar(\partial/\partial t)$ is equivalent to the energy operator $\widetilde{{\cal H}}$ in one and only one representation.

Now we need to determine the basic representation in order to calculate the EEVs.
It has been (explicitly or implicitly) supposed in preceding investigations \cite{Nietoold,Gol,Nieto,Fearing,Kupersztych} that the Dirac Hamiltonians and the Dirac wave functions
satisfy Eq. (\ref{EEV}). We will obtain a different result below. 

\section{Fundamental role of the Foldy-Wouthuysen
representation in determination of the energy expectation values}\label{Fundamental}

A determination of the basic representation results from: \emph{i)} an ascertainment of a classical limit of the
relativistic QM and \emph{ii)} a comparison of classical and quantum-mechanical Hamiltonians and equations of motion. The choice of the Dirac representation as a basic one \cite{Nietoold,Gol,Nieto,Fearing,Kupersztych} may by mostly motivated by the perfect covariance of the Dirac equation. On the other hand, the fundamental role of the FW representation in QM has become evident relatively recently.

It has been proven in Ref. \cite{JINRLett12} (with the extension of the Wentzel-Kramers-Brillouin method) that the transition to the classical limit of relativistic QM in the FW representation is obtained by the replacement of operators in the quantum-mechanical Hamiltonians and equations of motion with the respective classical quantities. This wonderful property shows that the relativistic quantum-mechanical equations for particles with
different spins should become very similar after the FW transformation. Thus, this transformation results in a unification of the relativistic QM.

Otherwise, investigations performed during last twenty years in the framework of the FW transformation in \emph{relativistic} QM (see Refs. \cite{TMP1995,relativistic,ChenChiou10,JMP,JMPcond,TMPFW} and references therein) have ascertained a strong resemblance between the 
Hamiltonians and equations of motion in the FW representation and the corresponding classical counterparts. It is important that such a resemblance covers all considered stationary and nonstationary problems in electrodynamics \cite{JMP,TMPFW,JETP2,RPJ,TMP2008,PhysRevDspinunit,PRDexact} and gravity \cite{gravity,PRD2007,OST,OSTgrav,Honnefscalar}. It holds true for relativistic particles with spins zero \cite{TMPFW,TMP2008}, one-half \cite{JMP,TMPFW,RPJ} and unity \cite{TMPFW,JETP2,PhysRevDspinunit,PRDexact} in arbitrary (generally, strong) time-dependent electromagnetic fields as well as for Dirac particles in arbitrary (generally, strong) time-independent \cite{gravity,PRD2007,OST} and time-dependent \cite{OSTgrav} gravitational fields and noninertial frames. A similar result has been recently obtained for spin-0 particles in 
gravitational fields and noninertial frames \cite{Honnefscalar}. It is instructive to mention that the quantum-mechanical description of single particles in strong external fields does not allow for specific effects of quantum field theory except for a phenomenological treatment of anomalous magnetic moments.

We can conclude that the above mentioned replacement of operators brings the relativistic quantum-mechanical FW Hamiltonians to the corresponding classical Hamiltonians. The considered properties cause relativistic QM in the FW representation to be analogous to nonrelativistic QM.

We need to comment on the relation between the operator $\bm r$ in the FW Hamiltonians and the radius-vector $\bm r$ in classical physics. The latter quantity corresponds to the Newton-Wigner position operator \cite{NW} (``mean position operator'' \cite{FW}) which is equal to $\bm r$ only in the FW representation. In the Dirac representation, this operator substantially differs from $\bm r$ and is given by a cumbersome formula \cite{FW}.

The \emph{operators} of canonical variables, $x^\mu$ and $p_\mu$, are equal to $x^\mu$ and $i\hbar(\partial/\partial x^\mu)$, respectively, in one and only one representation. The previous explanations definitely show that this is the FW representation. In classical physics, $p_0$ is equal to the Hamiltonian which defines the particle energy and is a function of $\bm r,\bm p, t$, and the spin $\bm s$.
In the FW representation, the operator $p_0=i\hbar(\partial/\partial t)$ should be equal to the Hamiltonian operator and should define the particle energy. As a result, the Hamiltonian operator is equal to the energy operator just in this representation:
\begin{equation} {\cal H}_{FW}=\widetilde{{\cal H}_{FW}}. \label{EEVdefi}
\end{equation} Therefore,
\begin{equation} E(t)=\int{\psi_{FW}^\dag(\bm r,t){\cal H}_{FW}(t)\psi_{FW}(\bm r,t)dV}. \label{EEVFW}
\end{equation}

In the Dirac representation, $x^\mu$ and $i\hbar(\partial/\partial x^\mu)$ [in particular, $i\hbar(\partial/\partial t)$] are \emph{not} the operators of canonical coordinates and momenta. In this representation, the determination of the EEVs should therefore be based on the general formulas (\ref{EEVprim}) and (\ref{EEVprimEq}). In these formulas, the operator $U$ is the operator of transformation \emph{from the FW to the Dirac representation}.

Thus, the nonequivalence of Hamiltonians in different representations does \emph{not} lead to the ambiguity of the EEVs.

Let us consider a spin-1/2 particle with anomalous magnetic and electric dipole moments in strong time-dependent electromagnetic fields as an example of the fundamental role of the FW
representation. In this case, the FW Hamiltonian has the form \cite{RPJ}
\begin{equation}\begin{array}{c}
{\cal H}_{FW}=\beta\epsilon'+e\Phi+\frac
   14\left\{\left(\frac{\mu_0m}{\epsilon'
   +m}+\mu'\right)\frac{1}{\epsilon'},\biggl(\bm\Sigma\!\cdot\![\bm\pi\!
\times\!\bm E]-\bm\Sigma\!\cdot\![\bm E\!\times\!\bm\pi]-\hbar\nabla\!
\cdot\!\bm E\biggr)\right\}\\ -\frac
12\left\{\left(\frac{\mu_0m}{\epsilon'}
+\mu'\right), \bm\Pi\!\cdot\!\bm B\right\}\\
+\beta\frac{\mu'}{4}\left\{\frac{1}{\epsilon'(\epsilon'+m)},
\biggl[(\bm{B}\!\cdot\!\bm\pi)(\bm{\Sigma}\!\cdot\!\bm\pi)+ (\bm{\Sigma}
\!\cdot\!\bm\pi)(\bm\pi\!\cdot\!\bm{B})+2\pi\hbar(\bm\pi\!\cdot\!\bm j+
\bm j\!\cdot\! \bm\pi)\biggr]\right\}\\
-d\bm\Pi\!\cdot\!\bm E
+\frac{d}{4}\left\{\frac{1}{\epsilon'(\epsilon'+m)},
\biggl[(\bm{E}\!\cdot\!\bm\pi)(\bm{\Pi}\!\cdot\!\bm\pi)+ (\bm{\Pi}
\!\cdot\!\bm\pi)(\bm\pi\!\cdot\!\bm{E})\biggr]\right\} \\-\frac
d4\left\{\frac{1}{\epsilon'},\biggl(\bm\Sigma\!\cdot\![\bm\pi\!
\times\!\bm B]-\bm\Sigma\!\cdot\![\bm
B\!\times\!\bm\pi]\biggr)\right\},
\end{array} \label{eq33new} \end{equation}
where $\bm\pi=\bm p-e\bm A\equiv-i\hbar\nabla-e\bm A$ is the kinetic momentum
operator, $\mu_0=e\hbar/(2m)$ and $\mu'=(g-2)e\hbar/(4m)$ are the Dirac and anomalous magnetic moments, $d$ is the electric dipole moment, $\epsilon'=\sqrt{m^2+\bm{\pi}^2}$, and $\bm j=(1/4\pi)\left(\nabla\times\bm B-\partial \bm E/\partial t\right)$
is the density of external electric current. To obtain the classical limit of the FW Hamiltonian, we set the Planck constant to zero ($\hbar\rightarrow0$) and substitute the classical quantities for the operators. As a result, we arrive at the equation
\begin{equation}\begin{array}{c}
H=\epsilon'+e\Phi+\bm s\cdot\bm\Omega,
\end{array} \label{eqclassnew} \end{equation} where $\epsilon'$ is the classical counterpart of the corresponding operator and $\bm\Omega$ is the angular velocity of spin precession:
\begin{equation}\begin{array}{c}
\bm\Omega=\frac{2}{\hbar}\left[\left(\frac{\mu_0m}
{\epsilon '+m}+\mu'\right)\frac{1}{\epsilon
'}\bm{\pi}\times\bm E-\left(
\frac{\mu_0m}{\epsilon '}+\mu'\right)\bm B
+\frac{\mu'}{\epsilon
'(\epsilon '+m)} \bm \pi(\bm \pi\cdot\bm B)
\right.\\ \left.
-d\bm E+\frac{d}{\epsilon '(\epsilon
'+m)}\bm \pi(\bm \pi\cdot\bm E)-
\frac{d}{\epsilon
'}\bm \pi\times\bm B\right].
\end{array} \label{eqclass} \end{equation}
In classical physics, the Hamiltonian and the angular velocity of spin precession \cite{GBMT} are defined by the same equations as Eqs. (\ref{eqclassnew}) and (\ref{eqclass}).

Now we can check the consequences of the assumption that the Dirac representation is the basic one. With this assumption, the difference between the energy operator and the FW Hamiltonian is given by
\begin{equation} \begin{array}{c} \widetilde{{\cal H}_{FW}}-{\cal H}_{FW}=-i\hbar\frac{\partial
U_{FW}}{\partial t}U_{FW}^{-1}. \end{array} \label{EEVprmEq}
\end{equation}
The right-hand side of this equation contains both even and odd terms. However, odd terms can be disregarded. Since the FW wave functions have only one nonzero spinor (upper and lower for states with positive and negative total energy, respectively \cite{JMPcond}), averaging the odd terms eliminates their contribution to the EEVs.

Partial derivatives with respect to time are hereinafter denoted by dots. The relativistic method of the FW transformation \cite{JMP,TMPFW} allows us to derive the following equation for the even part of $\widetilde{{\cal H}_{FW}}-{\cal H}_{FW}$:
\begin{equation}\begin{array}{c}
\widetilde{{\cal H}_{FW}}-{\cal H}_{FW}=\frac
   14\left\{\frac{\mu_0m}{\epsilon'(\epsilon'
   +m)},\biggl[\bm\Sigma\cdot\Bigl(\bm\pi
\times\dot{\bm A}-\dot{\bm A}\times\bm\pi
\Bigr)-\hbar\nabla
\cdot\dot{\bm A}\biggr]\right\}\\
+\beta\frac{\hbar}{8}\left\{\frac{1}{\epsilon'(\epsilon'+m)},
\biggl[\mu'\Bigl(\bm\pi\cdot\dot{\bm E}+\dot{\bm E}\cdot\bm\pi\Bigr)-d\Bigl(\bm\pi\cdot\dot{\bm B}+\dot{\bm B}\cdot\bm\pi\Bigr)\biggr]\right\}.
\end{array} \label{eqtnnew} \end{equation} Terms presented in this equation are exact.
Terms of the second and higher orders in $\hbar$ which do not relate to the contact interactions are not taken into account ($\mu_0,~\mu'$, and $d$ are proportional to $\hbar$).
An importance of terms presented in Eq. (\ref{eqtnnew}) for a derivation of the EEVs has been shown in Ref. \cite{Fearing}. In this work, the nonrelativistic approximation has been used.

Evidently, the energy operator corresponds to the classical Hamiltonian. Therefore, the assumption of the basic character of the Dirac representation \cite{Nietoold,Gol,Nieto,Fearing,Kupersztych} destroys the agreement between the relativistic QM and the classical physics.
The considered example confirms the fundamental role of the FW representation in relativistic QM, in particular, in the determination of the EEVs.

\section{Derivation of the energy expectation values in the Dirac representation}

Quantum-mechanical equations are usually solved in the Dirac representation. A derivation of the general equation for the EEVs in this representation is therefore rather important.
For this purpose, it is convenient to split the Dirac Hamiltonian into even and odd operators commuting and noncommuting with the operator
$\beta$, respectively:
\begin{equation} {\cal H}=\beta m+{\cal E}+{\cal
O}, ~~~\beta{\cal E}={\cal E}\beta, ~~~\beta{\cal O}=-{\cal
O}\beta. \label{eq3general} \end{equation} Even and odd operators are diagonal and off-diagonal in two spinors, respectively. To fulfill the FW transformation of the initial Hamiltonian (\ref{eq3general}), one uses \emph{a priori} information about commutation relations. Any commutator of the momentum and coordinate operators adds the factor $\hbar$, while a commutator of different Pauli (or Dirac) matrices does not affix such a factor. So one supposes that commutators like
$[{\cal O},{\cal E}]$ have the additional factor $\hbar$ as compared with the product of operators ${\cal O}{\cal E}$.
Since the Pauli matrices do not commute with each other, we assume that
multiple commutators of the form $[{\cal O},[{\cal O},\dots[{\cal O},{\cal E}]\dots]]$ add the factor $\hbar$ with respect to the operator
product ${\cal O}{\cal O}\dots{\cal O}{\cal E}$. This factor already appears due to the first commutation.
Since ${\cal O}^2$ is an even (block-diagonal) operator, the commutators of the forms $[{\cal O}^2,[{\cal
O},{\cal E}]]$, $[{\cal O}^2,[{\cal O}^2,{\cal E}]]$, and $[[{\cal
O},{\cal E}],{\cal E}]$ add the factor $\hbar^2$ as compared with the corresponding products of the operators. Contemporary methods of the relativistic FW transformation use an expansion in power series in the Planck constant \cite{relativistic,TMPFW}.

Equations (\ref{TrHam}) and (\ref{EEVprimEq}) show that the energy operator 
in the Dirac representation is defined by
\begin{equation} \widetilde{{\cal H}_{D}}={\cal H}_D+i\hbar\left(U\frac{\partial }{\partial t}U^{-1}-\frac{\partial }{\partial t}\right), ~~~ U=U_{FW}^{-1},
\label{TrHamDi}
\end{equation}
where $U$ and $U_{FW}$ are the transformation operators from the FW representation to the Dirac one and other way round, respectively.

Let us determine $\widetilde{{\cal H}_{D}}$ with allowance for terms proportional to the zeroth and first powers of $\hbar$. The relativistic FW transformation is fulfilled by iterative methods \cite{JMP,TMPFW} and the total transformation operator has the form $U_{FW}=\dots\cdot U_2U_1$. Since the first transformation performed with the operator $U_1$ eliminates the main odd terms, $1-U_2\sim\hbar$. With the given accuracy, $i\hbar U_2^{-1}(\partial/\partial t)U_2\approx i\hbar (\partial/\partial t)$. The transformation with the operator \cite{JMP} (see also Ref. \cite{TMPFW})
$$U_1=\frac{\epsilon+m+\beta{\cal
O}}{\sqrt{2\epsilon(\epsilon+m)}}$$
results in
\begin{equation} \widetilde{{\cal H}_{D}}={\cal H}_D+i\frac\hbar8\left\{\frac{1}{\epsilon(\epsilon
+m)},\left(\beta\{\epsilon,\dot{{\cal O}}\}+2\beta m\dot{{\cal O}}-\beta\{\dot{\epsilon},{\cal O}\}+[{\cal O},\dot{{\cal O}}]\right)\right\},
\label{TrHamfinal}
\end{equation} where $\epsilon=\sqrt{m^2+{\cal O}^2}$.

This general equation provides one with the possibility of calculating the EEVs with time-dependent Dirac Hamiltonians.

As an example, we can consider a spin-1/2 particle
in strong time-dependent electromagnetic fields. In this case, the Dirac Hamiltonian has the form (\ref{eq3general})
where
\begin{equation} {\cal E}=e\Phi-\mu'\bm
{\Pi}\cdot \bm{B}-d\bm {\Pi}\cdot \bm{E}, ~~~{\cal
O}=c\bm{\alpha}\cdot\bm{\pi}+i\mu'\bm{\gamma}\cdot\bm{
E}-id\bm{\gamma}\cdot\bm{B}. \label{DireqEDM} \end{equation}

The energy operator which defines the EEVs by averaging is given by
\begin{equation}\begin{array}{c} \widetilde{{\cal H}_{D}}={\cal H}_D+\frac{e\hbar}{8}\left\{\frac{1}{\epsilon'(\epsilon'
+m)},\left[-i\{\epsilon',\bm{\gamma}\cdot\dot{\bm A}\}-2im\bm{\gamma}\cdot\dot{\bm A}+\bm\Sigma\cdot(\bm\pi
\times\dot{\bm A}-\dot{\bm A}\times\bm\pi)\right]\right\}\\
+i\frac{e\hbar}{8}\left\{\frac{1}{{\epsilon'}^2(\epsilon'
+m)},\left[(\bm{\pi}\cdot\dot{\bm A})(\bm{\gamma}\cdot{\bm\pi})+(\bm{\gamma}\cdot{\bm\pi})(\dot{\bm A}\cdot\bm{\pi})\right]\right\}.
\end{array} \label{TrHamfnEEV}
\end{equation}

The contribution to the EEVs given by the two last terms in Eq. (\ref{TrHamfnEEV}) can be rather important. In a similar case, the importance of such a contribution has been shown in Ref. \cite{Fearing} with the use of the nonrelativistic approximation.

\section{Exact Foldy-Wouthuysen transformation of nonstationary Hamiltonians}

The even (block-diagonal) form of the final Hamiltonian was the only condition of transformation
used by Foldy and Wouthuysen \cite{FW}. However, this condition does
not define the FW Hamiltonian unambiguously. The additional condition eliminating this ambiguity has been proposed by Eriksen \cite{E} and substantiated by Eriksen and Kolsrud \cite{EK}. Additional substantiation of the Eriksen method has been given in Ref. \cite{VJ}.

The operator transforming the initial Hamiltonian to the FW representation can be presented in the exponential form:
\begin{equation} U_{FW}=\exp{(i{S})}.\label{Vvetott} \end{equation}
The transformation remains unique if the operator ${S}$ is odd and Hermitian \cite{E,EK}. This condition is equivalent to \cite{E,EK}
\begin{equation} \beta U_{FW}=U^\dag_{FW}\beta.\label{Erikcon} \end{equation}
We kept above the term ``Hermitian'' used in Refs. \cite{E,EK} while the operator ${S}$ should also be self-adjoint (see the beginning of Sec. \ref{unitary}).

Eriksen \cite{E} has found the operator satisfying Eq.
(\ref{Erikcon}) and therefore performing the exact FW transformation:
\begin{equation}
U_{E}=U_{FW}=\frac{1+\beta\lambda}{\sqrt{2+\beta\lambda+\lambda\beta}},
~~~ \lambda=\frac{{\cal H}}{({\cal H}^2)^{1/2}}, \label{eqXXI}
\end{equation}
where $\lambda$ is the sign operator. 
The denominator is an even operator and commutes with the numerator \cite{E,EK} (see also Ref. \cite{JMPcond}).

In Refs. \cite{E,EK,VJ}, only the stationary case was considered. However, we can extend the Eriksen method on the nonstationary case under discussion. The operator $\lambda$ is the sign operator even in this case: $\lambda\psi=\pm\psi$ [$\psi$ is the initial wave function defined by Eq. (\ref{eqCorSc})]. As a result, the operators $1+\beta\lambda$ and $U_{E}$ cause either a lower or a upper spinor to vanish for positive and negative energy states, respectively. In the nonstationary case, these operators can be time dependent. Since the operator $i\hbar(\partial/\partial t)$ in the FW representation (but not in the Dirac one) corresponds to $p_0$ in classical physics, the Dirac operator $\partial/(\partial t)$ corresponds to the following FW operator:
\begin{equation}
\left({\frac{\partial }{\partial t}}\right)_{FW}=U_{E}\frac{\partial }{\partial t}U_{E}^{-1}.
\label{eqXXnew}
\end{equation}
As \begin{equation} i\hbar\frac{\partial\psi_{FW}}{\partial t}={\cal
H}_{FW}\psi_{FW},
\label{eqCorScFW}
\end{equation}
the \emph{exact} FW Hamiltonian is equal to
\begin{equation} {\cal H}_{FW}=U_{E}\left({\cal H}-i\hbar\frac{\partial}{\partial
t}\right)U_{E}^{-1}+i\hbar\frac{\partial}{\partial t}.
\label{eqHamFW}
\end{equation}

While Eq. (\ref{eqHamFW}) solves the problem of the exact FW transformation in the nonstationary case, an explicit exact FW Hamiltonian can be obtained only in some special cases. In the general case, only an approximate expression for the FW Hamiltonian can be derived (see Ref. \cite{TMPFW}).

A sufficient condition for the exact FW transformation has been found in Refs. \cite{JMP,JINRLEriksen} for the stationary case. In the nonstationary case, it takes the form
\begin{equation} [{\cal F},{\cal
O}]=0, ~~~ {\cal F}={\cal E}-i\hbar\frac{\partial}{\partial
t}.
\label{eqexnFW}\end{equation}
When it is satisfied, the FW Hamiltonian is given by
\begin{equation}
{\cal H}_{FW}=\beta\epsilon+ {\cal E}, ~~~ \epsilon=\sqrt{m^2+{\cal
O}^2}.
\label{HmexnFW}\end{equation}

Possibilities of satisfying the condition (\ref{eqexnFW}) are very restricted. In particular, the operators $\partial/(\partial
t)$ and ${\cal O}$ do not commute for a spin-1/2 particle in nonstationary electromagnetic fields because $\dot{\cal O}\neq0$. Nevertheless, we can give an example of the exact FW transformation in the nonstationary case. Let us consider the Dirac particle in a nonstationarily rotating frame. The angular velocity of frame rotation, $\bm\omega(t)$, may arbitrarily depend on time. This frame is flat and its metric is given by
\begin{equation}\label{LT}
ds^2 = c^2dt^2 - (d\bm r+[\bm\omega(t)\times\bm r]dt)^2. 
\end{equation}
The corresponding Dirac Hamiltonian is equal to \cite{HN,OST}
\begin{equation}
{\cal H} = \beta m+\bm\alpha\cdot\bm p-\bm\omega(t)\cdot\left(\bm r\times\bm p+\frac{\hbar\bm\Sigma}{2}\right).
\label{Hamrf}\end{equation}

This Hamiltonian satisfies the condition (\ref{eqexnFW}) and its FW transformation is exact. The transformed Hamiltonian is given by
\begin{equation}
{\cal H}_{FW}=\beta\sqrt{m^2+\bm p^2}-\bm\omega(t)\cdot\left(\bm r\times\bm p+\frac{\hbar\bm\Sigma}{2}\right).
\label{HamFW}\end{equation}

This is the first example of the exact FW transformation in the nonstationary case. For a stationarily rotating frame ($\bm\omega=const$), the FW Hamiltonian has been derived in Ref. \cite{PRD2007}. The exact operator equation of spin motion is given by
\begin{equation}
\frac{d\bm\Sigma}{dt}=-\bm\omega(t)\times\bm\Sigma.
\label{OmegaFW}\end{equation}
Thus, the spin rotates with the instantaneous angular velocity $-\bm\omega(t)$. This conclusion fully agrees with classical gravity.

\section{Spin-1/2 particle in a plane monochromatic electromagnetic wave}

As an example demonstrating the validity of Eq. (\ref{EEVdefi}) and the invalidity of Eq. (\ref{EEVprmEq}), we can consider a spin-1/2 particle in a plane monochromatic electromagnetic wave. In this case, the conventional approach consists in
\begin{equation}
\Phi=0,~~~ \bm A=\frac{\bm E}{i\kappa}, ~~~ \bm E=\bm E_0\exp{[i(\bm\kappa\cdot\bm r-\omega' t)]}, ~~~ \bm B=\bm n\times\bm E, ~~~ \bm\kappa=\frac{\omega'}{c}\bm n, \label{planewave}\end{equation}
where $\bm n$ and $\omega'$ are the direction and the angular frequency of the wave. The corresponding Dirac equation admits an exact solution obtained by Volkov (see Ref. \cite{BLP}). The FW transformation is not exact but it ensures a high accuracy.

It has been mentioned in Sec. \ref{Fundamental} that averaging eliminates the contribution of odd terms in the operator $-i\hbar(\partial
U_{FW}/\partial t)U_{FW}^{-1}$ to the EEVs. The leading even term in this operator is proportional to $[{\cal O},\dot{\cal O}]$ and therefore contains the operator $\hbar\bm\Sigma$. As a result, it significantly affects the spin motion while its influence on the evolution of the momentum is rather weak.

The FW Hamiltonian of the particle is given by the general equation (\ref{eq33new}) where the fields are presented by Eq. (\ref{planewave}).
If we consent to the fundamental role of the FW representation in a determination of the EEVs, the classical limit of the operator of angular velocity of spin precession is presented by Eq. (\ref{eqclass}) and the spin motion fully corresponds to the Thomas-Bargmann-Michel-Telegdi \cite{Thomas,BMT} equation.
If the fundamental role of the Dirac representation in such a determination is assumed, the energy operator of the particle is equal to $\widetilde{{\cal H}_{FW}}$ and is defined by Eqs. (\ref{EEVprmEq}) and (\ref{eqtnnew}). In the considered case, $\dot{\bm A}=-c\bm E$. When all terms of the second order in $\hbar$ are disregarded, the above equations result in
\begin{equation}\begin{array}{c}
\widetilde{{\cal H}_{FW}}={\cal H}_{FW}-\frac
   14\left\{\frac{\mu_0m}{\epsilon'(\epsilon'
   +m)},\bm\Sigma\cdot\Bigl(\bm\pi
\times\bm E-\bm E\times\bm\pi
\Bigr)\right\}\\
=\beta\epsilon'+\frac
   {\mu'}{4}\left\{\frac{1}{\epsilon'},\bm\Sigma\cdot\biggl(\bm\pi
\times\bm E-\bm E\times\bm\pi\biggr)\right\} 
-\frac
12\left\{\left(\frac{\mu_0m}{\epsilon'}
+\mu'\right), \bm\Pi\!\cdot\!\bm B\right\} \\
+\beta\frac{\mu'}{4}\left\{\frac{1}{\epsilon'(\epsilon'+m)},
\biggl[(\bm{B}\cdot\bm\pi)(\bm{\Sigma}\cdot\bm\pi)+ (\bm{\Sigma}
\cdot\bm\pi)(\bm\pi\cdot\bm{B})\biggr]\right\}.
\end{array} \label{eqnwave} \end{equation}

The classical limit of the energy operator is the classical Hamiltonian.
In this limit, the angular velocity of spin precession corresponding to Eq. (\ref{eqnwave}) is equal to [see Eq. (\ref{eqclassnew})]
\begin{equation}\begin{array}{c}
\widetilde{\bm\Omega}=\bm\Omega-\frac{2}{\hbar}\cdot\frac{\mu_0m}
{\epsilon'(\epsilon'
   +m)}\bm{\pi}\times\bm E\\=\frac{2}{\hbar}\left[\frac{\mu'}{\epsilon
'}\bm{\pi}\times\bm E-\left(
\frac{\mu_0m}{\epsilon '}+\mu'\right)\bm B
+\frac{\mu'}{\epsilon
'(\epsilon '+m)} \bm \pi(\bm \pi\cdot\bm B)
\right],
\end{array} \label{eqtilda} \end{equation}
where $\bm\Omega$ is given by Eq. (\ref{eqclass}) (with $d=0$). The quantity $\widetilde{\bm\Omega}$ disagrees with the Thomas-Bargmann-Michel-Telegdi result.
This demonstrates that the supposition about the fundamental role of the Dirac representation in the determination of the EEVs is incorrect.

The EEVs in the Dirac representation are defined by Eqs. (\ref{TrHamfinal}) and (\ref{TrHamfnEEV}). With allowance for terms proportional to the zeroth and first powers of $\hbar$, they take the form
\begin{equation}\begin{array}{c} \widetilde{{\cal H}_{D}}={\cal H}_D+\frac{e\hbar}{8}\left\{\frac{1}{\epsilon'(\epsilon'
+m)},\left[i\{\epsilon',\bm{\gamma}\cdot\bm E\}+2im\bm{\gamma}\cdot\bm E-\bm\Sigma\cdot(\bm\pi
\times\bm E-\bm E\times\bm\pi)\right]\right\}\\
-i\frac{e\hbar}{8}\left\{\frac{1}{{\epsilon'}^2(\epsilon'
+m)},\left[(\bm{\pi}\cdot\bm E)(\bm{\gamma}\cdot{\bm\pi})+(\bm{\gamma}\cdot{\bm\pi})(\bm E\cdot\bm{\pi})\right]\right\}.
\end{array} \label{fnHamfnEEV}
\end{equation}

\section{Summary}

Thus, we confirm the result of the previous investigation \cite{Nieto} that the nonequivalence of Hamiltonians in different representations does not lead to an ambiguity of the EEVs. We show that the origin of this nonequivalence is a change of the form of the time derivative operator at a time-dependent unitary transformation. For a particle in nonstationary fields, the energy operator is equal to $U{\cal H}U^{-1}$ and does not coincide with the transformed Hamiltonian. Expectation values of the energy operator define the EEVs \cite{Nieto}.
However, it has been explicitly or implicitly supposed in Refs.
\cite{Nietoold,Gol,Nieto,Fearing,Kupersztych} that the basic representation in the time-dependent case is the Dirac one.
We prove that the comparatively recent developments of the theory of the relativistic FW
transformation lead to an alternative conclusion about the basic role of the FW representation. As an example of the importance of this problem, we have considered the spin-1/2 particle with anomalous magnetic and electric dipole moments in strong time-dependent electromagnetic fields. The
supposition that the Dirac representation is basic leads to a wrong description of the particle spin motion in this case and, in particular, in the case of a particle in a plane monochromatic electromagnetic wave.

This result is very natural. The operator $i\hbar\,(\partial/\partial x^i)~(i=1,2,3)$ in the Dirac representation does not correspond to the classical momentum $p_i$ and also the operator $x^i$ in this representation does not correspond to the classical coordinate. Therefore, the assumption that the operator $i\hbar\,(\partial/\partial t)$ in this representation is the energy operator and corresponds to the classical energy $p_0\equiv E$ postulates different properties of the spatial and temporal components of the operator $i\hbar\,(\partial/\partial x^\mu)~(\mu=0,1,2,3)$ and contradicts the relativistic invariance of the Dirac equation.

Since quantum-mechanical equations are usually solved in the Dirac representation, we have derived the general equation for the EEVs in this representation.
We have also found the sufficient condition of the exact FW transformation in the nonstationary case and have given the first example of such a transformation (the Dirac particle in a nonstationarily rotating frame).

\section*{Acknowledgements}

The author is grateful to M. Arminjon and V. P. Neznamov for helpful initial discussions.
The work was supported in part by the Belarusian Republican Foundation for
Fundamental Research (Grant No. $\Phi$14D-007) and by the Heisenberg-Landau program of the German Ministry for Science and Technology (BMBF).



\begin{thebibliography}{}

\bibitem{FW}
L.\,L. Foldy and S.\,A. Wouthuysen,  Phys. Rev. {\bf 78}, 29
(1950).

\bibitem{Messiah}
A. Messiah, \emph{Quantum Mechanics} (Wiley, New York,
1958), Vol. II, Chap. XX, Sec. 33.

\bibitem{Nietoold}
M. M. Nieto, Los Alamos Report No. LA-UR-76-2125 (unpublished).

\bibitem{Gol}
T. Goldman, Phys. Rew. D \textbf{15}, 1063 (1977).

\bibitem{Nieto}
M. M. Nieto, Phys. Rew. Lett. \textbf{38}, 1042 (1977).

\bibitem{Fearing}
S. Scherer, G. I. Poulis, H. W. Fearing, Nucl. Phys. A \textbf{570}, 686 (1994).

\bibitem{Arminjon}
M. Arminjon, Int. J. Theor. Phys. \textbf{52}, 4032 (2013);
arXiv:1302.5584; arXiv:1312.6707.

\bibitem{GorNeznPRD}
M. V. Gorbatenko and V. P. Neznamov, Phys. Rev. D \textbf{83},
105002 (2011).

\bibitem{GorNeznJdP}
M. V. Gorbatenko and V. P. Neznamov,
Ann. Phys. (Berlin), \textbf{526}, 195 (2014).

\bibitem{Kupersztych}
J. Kupersztych, Phys. Rew. Lett. \textbf{42}, 483 (1979).

\bibitem{footnote}
In this part of Sec. \ref{unitary}, we use denotations generally accepted in mathematics.

\bibitem{Reed-Simon}
M. Reed, B. Simon, \emph{Methods of Modern Mathematical Physics. I: Functional Analysis} (Academic Press, London, 1980), pp. 257-259.

\bibitem{Lusanna}
L. Lusanna and M. Pauri, Gen. Relativ. Gravit. \textbf{38}, 229 (2006).


\bibitem{JINRLett12}
A. J. Silenko, Pis'ma Zh. Fiz. Elem. Chast. Atom. Yadra \textbf{10},
144 (2013) [Phys. Part. Nucl. Lett. \textbf{10}, 91 (2013)].

\bibitem{TMP1995}
A. J. Silenko, Teor. Mat. Fiz. \textbf{105}, 46 (1995) [Theor. Math. Phys. \textbf{105}, 1224 (1995)].

\bibitem{relativistic}
A. J. Silenko, Phys. Rev. A \textbf{77}, 012116 (2008);
P. Gosselin, J. Hanssen, and H. Mohrbach, Phys. Rev. D \textbf{77}, 085008 (2008).

\bibitem{ChenChiou10}
T.-W. Chen and D.-W. Chiou,
Phys. Rev. A \textbf{89}, 032111 (2014); 
Phys. Rev. A \textbf{90}, 012112 (2014).

\bibitem{JMP}
A. J. Silenko, J. Math. Phys. {\bf 44}, 2952 (2003).

\bibitem{JMPcond}
V. P. Neznamov and A. J. Silenko, J. Math. Phys., \textbf{50},
122302 (2009).

\bibitem{TMPFW}
A. J. Silenko, 
Teor. Mat. Fiz. \textbf{176}, 189 (2013)
[Theor. Math. Phys. \textbf{176}, 987 
(2013)].

\bibitem{JETP2}
A. J. Silenko, JETP {\bf 96}, 775 
(2003); hep-th/0404074.

\bibitem{RPJ} A.\,J. Silenko, Izv. Vyssh. Uchebn. Zaved., Fiz. \textbf{48}, No. 8, 9 (2005)
[Russ. Phys. J. \textbf{48}, 788 (2005)]. 

\bibitem{TMP2008}
A. J. Silenko, Teor. Mat. Fiz. \textbf{156}, 398 (2008) [Theor.
Math. Phys. \textbf{156}, 1308 (2008)].

\bibitem{PhysRevDspinunit}
A. J. Silenko, 
Phys. Rev. D \textbf{87}, 073015 (2013).

\bibitem{PRDexact}
A. J. Silenko, 
Phys. Rev. D \textbf{89} 
121701(R) (2014).

\bibitem{gravity}
A. J. Silenko and O. V. Teryaev, 
Phys. Rev. D {\bf 71}, 064016 (2005); 
A. J. Silenko,
Acta Phys. Polon. B Proc. Suppl. {\bf 1}, 87 
(2008);
Yu. N. Obukhov, A. J. Silenko, and O. V. Teryaev,
Phys. Rev. D {\bf 84}, 024025 (2011). 

\bibitem{PRD2007}
A. J. Silenko and O. V. Teryaev, Phys. Rev. D {\bf 76}, 061101(R) (2007).

\bibitem{OST}
Yu. N. Obukhov, A. J. Silenko, and O. V. Teryaev,
Phys. Rev. D {\bf 80}, 064044 (2009).  

\bibitem{Honnefscalar}
A. J. Silenko, 
Phys. Rev. D \textbf{88}, 045004 (2013).

\bibitem{OSTgrav} Yu. N. Obukhov, A. J. Silenko, and O. V. Teryaev, 
Phys. Rev. D \textbf{88}, 084014 (2013).

\bibitem{NW}
T. D. Newton and E. P. Wigner, Rew. Mod. Phys. {\bf 21}, 400 (1949).

\bibitem{GBMT}
T. Fukuyama and A. J. Silenko, 
Int. J. Mod. Phys. A \textbf{28}, 
1350147 (2013).

\bibitem{E}
E. Eriksen, 
Phys. Rev. \textbf{111}, 1011 
(1958).

\bibitem{EK}
E. Eriksen and M. Korlsrud, Nuovo Cimento Suppl. \textbf{18},
1 (1960).

\bibitem{VJ}
 E. de Vries, J. E. Jonker,
Nucl. Phys. B \textbf{6}, 213 
(1968).

\bibitem{JINRLEriksen}
A. J. Silenko, Phys. Part. Nucl. Lett. \textbf{10}, 198 (2013).

\bibitem{HN}
F.~W.~Hehl and W.~T.~Ni, 
Phys. Rev. D {\bf 42}, 2045 (1990). 

\bibitem{BLP}
V. B. Berestetskii, E. M. Lifshitz, L. P. Pitaevskii, \emph{Quantum Electrodynamics}, 2nd ed. (Butterworth-Heinemann, Oxford, 1982).

\bibitem{Thomas}
L. H. Thomas, 
Nature (London) \textbf{117}, 514 (1926); Philos. Mag. {\bf 3}, 1 (1927).

\bibitem{BMT}
V. Bargmann, L. Michel and V. L. Telegdi, Phys. Rev. Lett. 
{\bf 2}, 435 (1959).

\end{thebibliography}
\end{document}